\documentstyle[12pt]{article}
\title{On edge states in superconductor with time inversion symmetry
breaking. }
\author{ G.E. Volovik\\
Low Temperature Laboratory\\
Helsinki University of Technology\\
P.O.Box 2200, FIN-02015 HUT, Finland\\
and\\
L.D. Landau Institute for Theoretical Physics, \\
Kosygin Str. 2, 117940 Moscow, Russia\\
}
\begin{document}

\maketitle
\vskip 2 truecm

\begin{abstract}
{Superconducting states with different internal topology are
discussed for the  layered high-$T_c$ materials. If the time inversion
symmetry is broken, the superconductivity is determined not only by the symmetry
of the superconducting state, but also by the topology of the ground state.
The latter is determined the integer-valued  momentum-space topological
invariant
$N$. The current carrying boundary between domains with different $N$ ($N_2\neq
N_1$) is considered. The current is produced by  fermion zero modes,
which number per one layer is $2(N_2-N_1)$.    }

\end{abstract} \vfill \eject

\section{Introduction}

Recently  a new phase transition in high temperature
superconductor was reported on
in Ni-doped Bi$_2$Sr$_2$CaCu$_2$0$_8$\cite{Movshovich}. It occurs at low
temperature, $T \sim 200~$mK and seems to correspond to the opening of the gap
in the superconductor which is  gapless (contains lines of nodes) above
transition. Since the gap usually  appears in superconductor with lines of nodes
if the time inversion (${\cal T}$) symmetry is violated, this
suggests that the spontaneous breaking  of the  ${\cal T}$-symmetry occurs in
the new transition. Such transition can be caused by interaction with magnetic
impurities.

Some evidence of  the additional symmetry breaking  with opening of
the gap in a pure Bi$_2$Sr$_2$CaCu$_2$0$_8$ in the presence of magnetic field
was reported in
\cite{KrishanaOng}. The
$d_{x^2-y^2}+id_{xy}$ complex order parameter was suggested in \cite{Laughlin}
to describe this experiment. The possibility of the broken
${\cal T}$-symmetry was discussed also within the grain boundaries and twin
boundaries \cite{Sigrist,Zhitomirsky}.

Since the broken
${\cal T}$-symmetry becomes popular now for the high temperature
superconductors, we discuss some consequences of the broken ${\cal T}$-symmetry
in D=2 spatial dimension relevant for layered superconductors. These
consequences are similar to that in thin films of superfluid
$^3$He-A and other 2D systems where
${\cal T}$-symmetry is  broken. They are described in terms of the
integer-valued topological invariant of the ground state
\cite{Volovik1989,Yakovenko1990,Yakovenko1991,Volovik1992,Exotic}, which gives
rise to chiral edge states on the boundary of superconductor and in some cases
to quantization of Hall conductivity. We dicuss here this topological invariant
for the layered superconductors.

\section{Internal topological invariant.}

In the BCS model of the high-$T_c$ superconductor, the quasiparticle
spectrum is obtained from the Bogoliubov-Nambu Hamiltonian
$${\cal H}= \tau_3
\epsilon ({\bf k}) +  \tau_1
\Delta_1 ({\bf k}) +\tau_2
\Delta_2 ({\bf k})\equiv  \vec\tau \cdot\vec m({\bf k})~~,
\eqno(2.1)$$
where $\tau_i$ are the $2\times 2$ Pauli matrices in particle-hole space;  ${\bf
k}$ is the 2D vector of the in-plane linear momentum.
In the simplest
model
$$ \epsilon ({\bf k})={{\bf k}^2-k_F^2\over 2m}~~,~~  \Delta_1 ({\bf k})=
d_1(n_x^2-n_y^2)+s_1 ~~,~~ \Delta_2 ({\bf k})=
2d_2n_xn_y +s_2 ~~,\eqno(2.2)$$
where $\hat n=(n_x,n_y)= {\bf k}/|{\bf k}|$.
The gap function $\Delta_1 ({\bf k})$
describes the original $d_{x^2-y^2}+s$ state of the high-$T_c$ superconductor,
where the admixture of the $s$ state comes from the orthorhombic deviation
of the
crystal symmetry from the tetragonal. It is believed that $s_1$ is
less than $ d_1$, so the gap nodes do not disappear in the mixture, but are
shifted from their symmetric positions in the tetragonal crystal.

The function $\Delta_2
({\bf k})$ appears in the state with broken ${\cal T}$-symmetry. The ${\cal
T}$-symmetry can be broken in two ways: with conservation of the symmetry
${\cal I}$ with respect to reflection in $\left<100\right>$ crystal axis (cf.
\cite{Annett}), or with conservation of the combined symmetry ${\cal IT}$.  In
the  ${\cal I}$-symmetric state one has $d_2=0$, while in the  ${\cal
IT}$-symmetric state $s_2=0$. However, we consider here the more general case,
when both
${\cal I}$ and ${\cal IT}$ are broken and one has a mixture of
$is$ and $id$ components. Our goal is to show that the properties of
the system at $T=0$ are determined by the
ground state topology rather than by symmetry.

In the broken
${\cal T}$-symmetry state the gap in the energy spectrum should  appear
$$\Delta={\rm min}~ |E({\bf k})|\neq 0~~,~~E^2({\bf k})=m^2({\bf
k})= \epsilon^2 ({\bf k})+\Delta_1^2({\bf k}) +\Delta_2^2({\bf
k})~~.\eqno(2.3)$$
There are exceptions from this rule: at some values of the  parameters
satisfying the condition
$$\left({s_1\over d_1}\right)^2+\left({s_2\over d_2}\right)^2=1~~
\eqno(2.4)$$  the gap $\Delta$ becomes zero at some points ${\bf k}$. This
equation thus determines the surfaces in the phase space of the  parameters
$s_1,d_1,s_2,d_2$, which separate the regions of the gapped superconducting
states. On these surfaces the superconductivity is gapless.

The gapped superconducting
states on different sides of the surface of gapless superconductivity have the
same symmetry. But these states are different in terms of the internal topology,
determined by the topological invariant of the ground state \cite{Exotic}:
$$N={1\over 4\pi} \int d^2k ~ \hat m\cdot \left( {\partial\hat m \over
\partial k_x}\times  {\partial\hat m \over \partial
k_y}\right) ~~,~~\hat m={\vec m\over |\vec m|} \eqno(2.5)$$
In our simple model this invariant is
$$N=0~~,~~{\rm if}~~~~\left({s_2\over d_2}\right)^2+\left({s_1\over
d_1}\right)^2 >1~~
\eqno(2.6)$$
and
$$N= 2 ~{\rm sign} ~(d_1d_2)~~,~~{\rm if}~~~~\left({s_2\over
d_2}\right)^2+\left({s_1\over d_1}\right)^2<1~~.
\eqno(2.7)$$
The sign of $N$ is determined by the relative sign of   real and imaginary
components of the $d$-wave order parameter.  The transition between the states
with different integer-valued invariant $N$  occurs via the gapless intermediate
state (see similar phenomenon in the thin films of $^3$He-A, where the ${\cal
T}$-symmetry is also broken \cite{Exotic}).

In general the order parameter is more complicated, moreover the  system of
interacting fermions cannot be described by the effective Hamiltonian of
the type
(2.1). The topological invariant of the ground state nevertheless exists, but it
is expressed in terms of the  Green's functions \cite{Volovik1989,Exotic}.

\section{Chiral edge states in  domain walls.}

Because of the 2-fold degeneracy of the ${\cal T}$-symmetry broken
states, there can be  domain walls:  surfaces in  the real
space separating the domains with opposite parity. If
$N\neq 0$, the degenerate superconducting states have opposite value of $N$.
Because of the jump of $N$ across the domain wall, such wall must contain the
fermion zero modes. These are the gapless branches $E(k_\parallel)$, where
$k_\parallel$ is the linear momentum along the wall, they cross zero
energy when $k_\parallel$ varies. These fermion zero modes correspond  to the
chiral gapless edge states in the Quantum Hall Effect (see
\cite{Volovik1992} for the $p$-wave pairing in $^3$He-A film and
\cite{Laughlin} for the pure
$d$-wave case with
$s_1=s_2=0$).
Close to zero energy the spectrum of the $a$-th mode is linear:
$$E_a(k_\parallel)=c_a (k_\parallel -k_a)~~
\eqno(3.1)$$

There is an index theorem which determines the  algebraic number
$\nu$  of the fermion zero modes, ie the number of modes crossing zero with
positive slope, minus the number of modes with negative
slope,
$$\nu=\sum_a {\rm sign}~c_a ~~
\eqno(3.1)$$
According to index theorem the number $\nu$ per one $CuO_2$ layer for a wall
separating the superconducting states with
$N=N_1$  and $N=N_2$  is
$$\nu=2(N_2-N_1) ~~.
\eqno(3.3)$$
If one considers the boundary between  superconductor and insulator, the
invariant
$N$ on the insulating side should be put zero.

The fermion zero modes in the domain wall have the same origin as the
the fermion zero modes in spectrum of the Caroli-de Gennes-Matricon bound states
in the vortex core, where the
${\cal T}$-symmetry is also broken \cite{Caroli}. In the latter
case the chiral fermions are orbiting around the vortex axis, which corresponds
to the motion along the  closed domain boundary. For the circular domain
 wall of radius $R$ the edge state have an angular momentum
$Q=k_\parallel R$, and from Eq.(3.1) one obtains the spectrum of the
low-energy bound states in terms of the angular momentum $Q$:
$$E_a(Q)=\omega_a (Q -Q_a)~~
\eqno(3.4)$$
Here $\omega_a=c_a/R$ is the angular velocity of the rotation along the
closed trajectory. This equation represents  the general spectrum of fermions
bound to the vortex core
\cite{Kopnin,VolovikOrbitalMomentum,KopninVolovik1997}.  The number of the
fermion zero modes in the vortex core, the branches $E_a(Q)$ which as
functions of $Q$ cross zero energy, is also determined by the index theorem
\cite{Q-modes-Index}:
$$\nu=\sum_a {\rm sign}~\omega_a =2N~~,
\eqno(3.5)$$
but now $N$ is the winding number of the vortex.

\section{Current carrying by edge states.}

The breaking of the ${\cal
T}$-symmetry leads to the ground-state current carried by the occupied negative
energy states in the domain wall. The topological characteristics of
the fermionic charge (current) accumulated by the general texture in $^3$He-A,
where  ${\cal
T}$-symmetry is broken,   was discussed in
\cite{StoneGaitan}. In our case the  current is concentrated within the
domain wall and this current along the closed wall leads to the angular momentum
of the domain.  The magnitude of the angular momentum can vary because of the
fermionic charge accumulated by the superconducting state. This accumulation
occurs due to spectral flow in the fermion zero modes (see
\cite{VolovikOrbitalMomentum}).

In the superconducting state is axisymmetric and has no additional breaking
of spatial parity, the angular momentum of superconductor  per electron
is quantized  at $T=0$:
$L_z={1\over 2}\hbar N$. This momentum does not depend on details of the gap
structure and is determined only by the topological invariant $N$. In our case
simple model the axisymmetric state with nonzero $N=2$ occurs if $s_1=s_2=0$ and
$d_1=d_2$. However when one deforms  the superconducting
state from the axisymmetric  to the relevant one, the momentum
$L_z$ will is substantially modified by the deformation, if the spectral flow
takes place during such a deformation
\cite{VolovikOrbitalMomentum}. But typically $L_z$ remains to be of the same
order.

One can estimate the change of the edge current under the deformation of
superconducting state from the most symmetric one. The mass current along the
domain wall changes in the following way:
$$\Delta J_M={e\over 8\pi \hbar} \sum_a k_a^2 ~{\rm sign} ~c_a~~.
\eqno(4.1)$$
This gives the following change of the electric current
$$\Delta J_e={e\over 8\pi \hbar}
\sum_a {k_a^2\over m} ~{\rm sign} ~c_a ~~.
\eqno(4.2)$$

The Eq.(4.1) is obtained from the following consideration. The change in the
current is caused by the spectral flow which takes place during the change
of the parameters of the system, $(s_1,s_2,d_1,d_2)$. If
$\delta n_a$ is the number of levels at $a$-th branch, which cross  zero energy
under the deformation of the state, the variation of the current is
$$\delta J_M= \sum_a k_a \delta n_a ~~,
\eqno(4.3)$$
since each level, when it crosses zero, carries the momentum $k_a$ with it.
On the other hand the variation of the number of levels is the variation of the
momentum divided by the phase space volume $2\pi \hbar$:
$$\delta n_a= {\delta  k_a
\over 2\pi \hbar}~{\rm sign} ~c_a ~~.
\eqno(4.4)$$
This gives
$$\delta J_M= {1  \over 8\pi \hbar} \sum_a   \delta (k_a^2 ) ~{\rm sign} ~c_a
~~.
\eqno(4.5)$$
An extra $1/2$ is introduced to compensate the duplication of degrees of
freedom, when both particle and hole states are considered in the Bogoliubov
description in Eq.(2.1). The Eq.(4.5) leads to Eq.(4.1).

In some special cases one has $k_a=0$ and the change in the edge
current becomes zero. This may happen if the spatial parity  ${\cal
P}$  is not violated and for special orientations
of the domain wall.  In general case the momentum, for which the energy is zero,
has a finite value, which is typically  of order
$k_F$.  There are can be special orientations of the wall for which
$k_a=\pm k_F$. In this case one obtains the quantized response of the current to
the chemical potential
$\mu$:
$\delta J_e= {e  \nu\over 4\pi \hbar}   \delta \mu$ \cite{Volovik1992}.
However this is not a general result.  In general $\Delta J_e \sim  e \nu
k_F^2 /
4\pi \hbar m  $ if
$\nu=\sum_a {\rm sign}~c_a\neq 0$.

\section{Conclusion.}

We described  different classes of superconductivity in $CuO_2$ planes with
broken time inversion symmetry. The superconducting states may have  the same
symmetry but differ by the topological properties.  The superconducting states
are described by the integer-valued momentum-space topological invariant $N$.
The boundary separating  domains with different  $N$ ($N_2\neq N_1$) contains
fermion zero modes, which number is determined by $N_2-N_1$. These fermions are
current carrying  and produce nonzero current along the
domain wall in the ground states if $N_2\neq N_1$.

The magnitude of the edge current is usually  of order $ e  k_F^2/ 4\pi \hbar
m $. It is large compared to the magnitude obtained
 by Laughlin (see  Eq.(21) of \cite{Laughlin}), who considered the pure
$d$-wave case, which has $N=2$. Laughlin's result can be obtained if one
takes $k_a^2/2 m= \Delta$ in Eq.(4.2), where
$\Delta$ is the gap in the spectrum in Eq.(2.3). However there is no reason for
such identification.

The domain wall fermions are also important in particle physics. For example
the chiral fermions in our 3+1 dimensions can be reproduced by zero modes
bound to domain wall in 4+1 dimensions \cite{KaplanSchmaltz}.

 This work was
supported by the Russian Foundation for Fundamental Sciences, Grant No.
96-02-16072.

\end{document}